\documentclass[conference]{IEEEtran}
\IEEEoverridecommandlockouts
\usepackage{cite}
\usepackage{amsmath,amssymb,amsfonts}
\usepackage{algorithmic}
\usepackage{graphicx}
\usepackage{textcomp}
\usepackage{xcolor}
\usepackage{url}
\usepackage{comment}

\def\BibTeX{{\rm B\kern-.05em{\sc i\kern-.025em b}\kern-.08em
    T\kern-.1667em\lower.7ex\hbox{E}\kern-.125emX}}
\begin{document}

\title{Rational Economic Behaviours in the Bitcoin Lightning Network}

\author{
    \IEEEauthorblockN{Andrea Carotti}
    \IEEEauthorblockA{
        acarot2@uic.edu\\
        University of Illinois at Chicago \\
        Chicago, Illinois, USA
    }
    \and
    \IEEEauthorblockN{Cosimo Sguanci}
    \IEEEauthorblockA{
        cosimo.sguanci@mail.polimi.it\\
        Politecnico di Milano \\
        Milan, Italy
    }
    \and
    \IEEEauthorblockN{Anastasios Sidiropoulos}
    \IEEEauthorblockA{
        sidiropo@uic.edu\\
        University of Illinois at Chicago \\
        Chicago, Illinois, USA
        }
    }

\maketitle

\begin{abstract}

The Bitcoin Lightning Network (LN) is designed to improve the scalability of blockchain systems by using off-chain payment paths to settle transactions in a faster, cheaper, and more private manner.
This work aims to empirically study LN’s fee revenue for network participants. Under realistic assumptions on payment amounts, routing algorithms and traffic distribution, we analyze the economic returns of the network's largest routing nodes which currently hold the network together, and assess whether the centralizing tendency is incentive-compatible from an economic viewpoint. 
Moreover, since recent literature has proved that participation is economically irrational for the majority of large nodes, we evaluate the long-term impact on the network topology when participants start behaving rationally.

\end{abstract}


\section{Introduction}
Payment channel networks, such as the Lightning Network (LN) \cite{poon2016bitcoin}, are designed to enhance the scalability, speed and privacy of blockchain transactions by shifting the bulk of transactions off the main blockchain. In this setup, nodes open bilateral payment channels by depositing funds into a shared multi-signature address, allowing them to conduct off-chain transactions without interacting with the blockchain. These channels link together to form a network, enabling payments to be routed through intermediary nodes that can earn a fee in return for their service. As a consequence, there is an incentive for nodes to lock up funds to provide the payment routing infrastructure.\\
The LN currently exhibits a high degree of centralization, with a few nodes holding most of the network liquidity in their channels. Considering the high unreliability of LN payments, to improve the payment success rate, routing algorithms in client implementations incorporated node capacity and node availability to compute the path from sender to receiver \cite{lnd_pathfinding}. As a consequence, the question arises whether these large hubs have a disproportionate advantage in terms of fee revenue \cite{issue_centralization} to maintain their status of central nodes or if they should split favoring decentralization. In this work we explore whether there is empirical proof of such centrality convenience, and if this behaviour is affected by changing the distribution assumption on payments and by changing the routing algorithm. Therefore, we propose strategy candidates from the perspectives of routing nodes to improve their current fee revenue, following rationality in the choice of channels and fees.\\
We study the global long-term impact of rational behaviour on the average fee in LN under the assumption of a dynamic framework, which we model as a repeated game. In our model, at each iteration the network graph is updated, allowing nodes to change their strategies. 
We consider different sets of strategic behaviour for the nodes: changing their fees, chancing their peers by re-allocating capital, and a combination of both.
When nodes do not change their peers, but only channel fees, our research predicts a long-term growth in mean LN fees, confirming the thesis that at the current state of the network channels are underpriced. 
Conversely, when peer modification is possible for nodes, the average transaction fee converges to lower values than the current ones.
The main contributions of our work can be summarized as follows:

\begin{itemize}
    \item We study the earnings for high-capacity nodes and assess their revenue advantage as a function of the routing algorithm.
    \item We illustrate simple strategies for nodes to increase fee revenue from routing by replicating other nodes.
    \item We analyze how network topology and fees change when nodes dynamically adapt their fees and channels.
\end{itemize}

The next sections are structured as follows. Section \ref{section:preliminaries} provides a detailed description of the model and the notation necessary for our research. In Section \ref{section:empirical_analysis} we empirically analyze fee revenue for a subset of the nodes in the network and we evaluate the \textit{``rich-gets-richer"} aspect of the LN \cite{barabasi1999emergence}. In Section \ref{section:nodes_rational_strategies} we analyze selfish strategies for increased revenue for larger nodes when they replicate nodes currently adopting a better strategy. Finally, in section \ref{section:long_term_impact_random_strategy} we analyze the dynamic evolution of the network when nodes adapt to the new network state in order to maximize their profit. Section \ref{section:related_work} gives an overview of related work, and Section \ref{section:conclusion} concludes the paper.

\section{Preliminaries}
\label{section:preliminaries}

From a snapshot taken with the LND client implementation on January 19, 2023, the LN consisted of more than 14000 nodes, and around 70900 payment channels with a combined capacity of more than 5179 BTC. From the perspective of network topology, it has been observed that it follows a power-law degree distribution \cite{seres2020topological, martinazzi2020evolving}. Specifically, the fraction of nodes \(P(k)\) having a degree
\(k\) is described as \(P(k) \sim k^{-\alpha}
\) with \(\alpha\) typically ranging
between 2 and 3 \cite{clauset2009power, wang2022can}.

\subsection{Network Model}

The LN can be defined as a multigraph, where each bi-directional payment channel is represented as a pair of edges between two nodes with opposite directions. The graph can be defined as \(G = (V, E)\), where the vertex set \(V\) constitutes the nodes and the multiset \(E\) the payment channels.
There is a function \(b\) that assigns a \emph{balance} to each edge \(e\) in the graph.
\[
  b: E \xrightarrow{} \mathbb{N}_0
\]

Every bidirectional channel is represented by two directed edges to separately store the individual capacities and channel policies of both channel endpoints. The balance in a channel is constrained by the channel capacity. The capacity is a  publicly known function \( c : E \rightarrow \mathbb{N}
\) that assigns a capacity to every edge of the network. For every edge \( e = (u,v) \) we denote \( e_u := (e, u) \) as the first participant of the channel and \(e_v = (e,v) \) as the second participant, while \( c(e) \) is the total capacity. The capacity of every channel \(e = (u, v) \) is privately split into the local balances with the balance function such that  \(b(e_u) + b(e_v) = c(e)\) \cite{pickhardt2020imbalance}.

To reduce the number of required channels, the Lightning Network offers multi-hop routing, which enables sending payments to non-adjacent nodes in the network. In this case, the payment is routed over intermediate nodes along some payment path. Because the sending node only knows the exact amount of liquidity in its own outgoing and inbound channels, it has to guess liquidity in the rest of the network.

\subsection{Payment distributions and demand}

For privacy reasons, LN does not publicly disclose the traffic distribution across the network. So, to obtain realistic empirical results we resort of using a simulator under certain assumptions about the distribution of payments.
Generally speaking, due to the nature of LN, we expect a large number of transactions for relatively small amounts and fewer transactions for larger amounts. 
For what concerns how payments are distributed, most of the previous empirical studies on LN fees \cite{lange2021impact, beres2021cryptoeconomic} assumed a uniform demand of payments across the network. %
Since larger nodes are likely to be involved in more payments (e.g.~large merchants, exchanges, and so on),
we also consider payment distributions that depend on node capacity.
We define the neighbor function as \(n : V \rightarrow 2^{E}\) that assigns to every node a set of all the channels it is part of, and we consider for simplification the neighborhood function \(U := n(u)\). we also consider the case where, for each node \(u \in V\), the demand of \(u\) is proportional to some function \(f(\tau(u))\).
We define \(\tau(u) := \sum_{e \in U} b(e_u) \), with the following mild statistical assumption:

$$\mathbf{E}[b(e_u)] = \frac{c(e_u)}{2}$$

Therefore, we can say that $\sum_{e \in U} \mathbf{E}[b(e_u)] = \sum_{e \in U} \frac{c(e_u)}{2}$. This assumption is made due to the inherent lack of knowledge on channel balances as expected by the protocol.
We define the total demand of a node $u \in V$ as:

\[
d(u) = \sum_{u \in V} f(\tau(u)) \cdot f(\tau(v)) = 
\]
\[
 =f(\tau(v)) \cdot \sum_{u \in V} f(\tau(u) = f(\tau(u)) \cdot C
\]

Now, depending on the function $f$ chosen, the probability of picking a node \(u\) as a sender changes.

\subsubsection{Uniform}
The Uniform payment distribution is the simplest. The probability that a node \(u \in V\) sends a payment is \(P(u) = \frac{1}{n}\) where \(n = |V|\).
\subsubsection{Proportional to capacity}
Another way to choose the sender (and the receiver) of a payment is through the demand of the node \(u\). In our framework, we tested \(f(x) = x\). In this case, the probability of picking \(u\) is defined as:
\[
P(u) = \frac{d(u)}{\sum_{v \in V} d(v)} = \frac{f(\tau(u))}{\sum_{v \in V} f(\tau(v))} 
\]

\subsection{LN clients and their routing algorithms}

The LN uses source routing, so the path chosen is implementation-dependent, meaning that it is the sender that decides the route to follow to reach the destination. The three main implementations to be found in LN are
\verb|LND| (87\% of nodes), \verb|C-lightning| (11\% of nodes), and \verb|Eclair|  (2\% of nodes) \cite{lnd_pathfinding}.\\
Each of those determines the best route by utilizing a customized variant of Dijkstra's shortest path algorithm \cite{dijkstra2022note}. In our analysis, we will use both the standard version of the algorithm that uses as edge weight the total fee paid by the sender, and some modified version of it that also takes into account channel capacity, as done by the state of the art implementations \cite{andreescu2021comparing, kotzer2023braess}.\\
In the LN, transaction fees can be computed because each edge stores routing fee policies. There is a fixed fee denoted as \textit{base fee} that is charged each time a payment is routed through the channel. This can be expressed with a function \(f_{base}: E \rightarrow \mathbb{N}_0 \). There is also a \textit{fee rate}, that is a percentage fee charged on the value of the transaction amount that is sent through that node, and can be expressed with \(f_{ppm}: E \rightarrow \mathbb{N}_0 \).\\
Both fees are initially established at the time of channel creation. However, these fees are not static; they can be dynamically adjusted by the user at any subsequent point in time, forming a fee market for payment routing. Given three nodes \(u\), \(v\) and \(w\) such that there is an open channel between \(u\) and \(v\) and between \(v\), and \(w\), the profit \(p\) of node \(v\) for forwarding a transaction from \(u\) to \(w\), of total amount \(amt\), is computed as: 

\[
p(v) = f_{base}(v, u) + f_{ppm}(v,u) \cdot amt \cdot 10^{-3}
\]
Note that $v$ earns a fee only on its outgoing edge involved in the payment path.
The amount \(amt\) is expressed in satoshis, while \(f_{base}\) and \(f_{ppm}\) in millisatoshis.

The standard version of Dijkstra takes only into account the fees to be paid to the routing node for shortest path computation. In line with current client implementations, the edge weight \(w(e)\) can be adjusted to also take into account the capacity.
\[
w(e) = fee_e + \alpha / c_e \tag{1}
\]
Where \(fee_e\) is the fee paid to route through edge \(e\) and \(c_e\) is the channel capacity of \(e\), both values should be normalized in some interval. In our analysis, we applied \textit{min max normalization}, considering only fees between 0 and 20,000 msats and capacities between 1 and the maximum channel capacity from the snapshot (500,000,000 sats).

The choice of $\alpha$ corresponds to a trade-off between cost and reliability: larger values for $\alpha$ increases the cost of routing, but is likely to result in higher payment success probability.
As illustrated in the following sections, this phenomenon can be exploited by nodes of high total capacity to increase their revenue.
The above weight is used in substitution of the Pickhardt routing algorithm \cite{pickhardt2021optimally}, according to which to obtain the most reliable path that takes into account fee and capacity, we need to use the most probable multi-part split for delivering the amount \(k\). This solution was also used in our analysis, but since it runs in \( O(log(k)·(m^{2} + mn)·log(n)) \) where \(n\) is the number of nodes, \(m\) is the number of edges, and \(k\) the amount to be delivered, computing all-source-shortest path with weight (1) makes the simulation more efficient compared to the Pickhardt implemented heuristic.

\section{Empirical analysis}
\label{section:empirical_analysis}

In the following, we empirically evaluate the profit of nodes with respect to their locked liquidity in the network. To do so we define the \emph{ratio} of a node \(u \in U\) as
\[
R(u) = \frac{\sum_{i=1}^{d} p_i(u)}{\tau(u)},
\]
where \(p(u)\) is the fee earned by node \(u\) from routing a transaction \(i\) in a simulation, and \(d\) is the number of routed payments of \(u\).
The following results can only be obtained through simulation as LN protocol does not allow users to know the traffic of other nodes. Another detail of our model is that we are not considering failed payments. So the results only show successful payments. Whenever a payment fails we simulate a new payment. The third aspect is that all our simulations use a fixed amount of 10,000 sats, which is currently less than 5\$. This is close to the median amount (24.4k sats) routed by some of the largest routing nodes that have made their statistics publicly available \cite{river_report_1}. Another key assumption in our study is the exclusion of channel balances from consideration, except in scenarios where we use the Pickhardt LN simulator. In these cases, the probability of a failed payment for 10,000 sats is less than 1\%. Since we are simulating small payments, it is unlikely that channels get exhausted \cite{rohrer2019discharged} as they are initialized at $\frac{c(e_u)}{2}$ and the median capacity of the subset of the network considered is 6,000,000 sats. For this reason, and for simplicity in our analysis, we ignore the issue of channel rebalancing.

\subsection{Experimental Methodology}

As a basis for the empirical analysis in the same fashion as \cite{lange2021impact, conoscenti2019hubs}, we developed a time-discrete event simulator that implements the network multi-graph model (cf. Section II-A) and allows payment simulation, as well as 
dynamic changes in the network that can either be the opening of channels with new nodes and changing channel fees. We also extended Pickhardt's publicly available simulator\footnote{The code implementing the simulations and the experiments can be found at the following URL: \url{https://drive.google.com/drive/folders/1i5EnnGNDzK3FeGj4_D_6-s4ucQ_HcNKk?usp=sharing} } to support multiple payment simulations according to given distributions.\\
Our simulator initially reads the network graph from a snapshot of the LN and simulates path finding through a weight-based route selection algorithm. Pickhardt's simulator instead implements the routing algorithm in these steps: it first computes the path probability on the \textit{uncertainty network } \cite{pickhardt2021optimally}, then uses a heuristic that minimizes the following cost function:
\[
C(f) =  
\sum_{e \in E} -\log \left( \frac{c_e + 1 - amt}{c_e + 1} \right) + \mu \cdot amt \cdot fee_e
\]

\(c_e\) is the channel capacity, \(amt\) is the amount to send, \(\mu\) is a constant set by the user, and \(fee_e\) is the fee to pay to route through edge \(e \in E\).\\
Since the analysis is about routing profits of service providers and routing nodes, we considered only the 1000 nodes with the highest capacity. Formally, given \(G = (V,E)\) we consider the set \(U = \{u_1, ..., u_{1000}\}\) such that \( \forall u \in U, \forall v \in V \setminus U : \tau(u) > \tau(v) \).\\ 

For this first analysis, we tested the routing algorithms described in Section II-C. The experiments assumed medium payments of 10,000 sats. We then evaluated for different values of \(\alpha\) and \(\mu\) considering our greedy routing algorithm that employs (1) as weight and Pickhardt routing algorithm respectively. The last aspect we highlight is that in this first set of results, no upper threshold was imposed on the fee to pay for sending a payment. This implies that, while in a practical scenario a user would avoid paying above a certain threshold, in this analysis, the focus is solely on the fee incurred, assuming the existence of a feasible path leading to the intended destination.

\begin{figure*}[t]
    \centering
    \includegraphics[width=\textwidth]{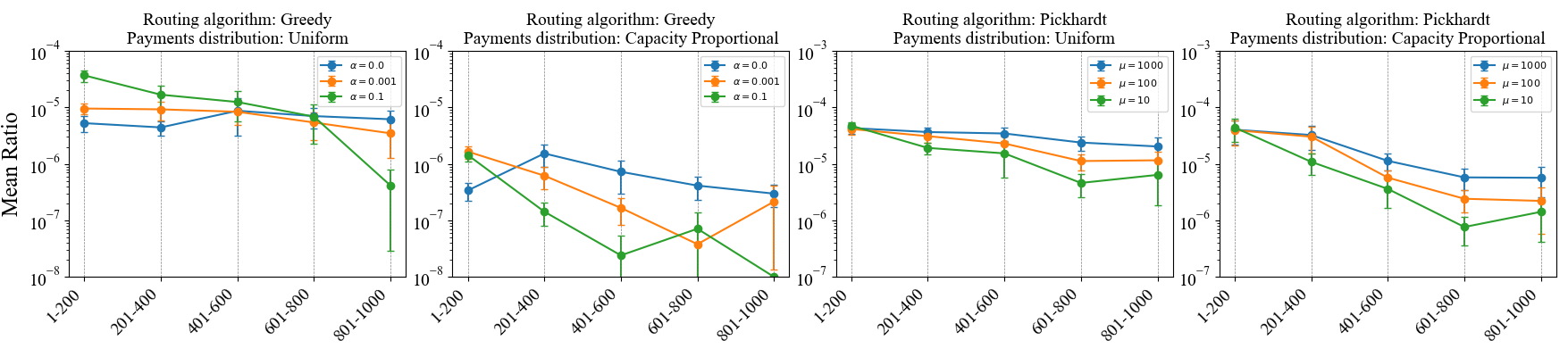}
    \caption{
        Mean Ratio of nodes for buckets of 200 nodes grouped by decreasing capacity using our greedy routing algorithm and Pickhardt routing algorithm. Results are shown with uniform and capacity-proportional distribution of payments in a simulation of 10,000 payments of 10,000 sats.
    }
    \label{fig:mean_ratio}
\end{figure*}

\subsection{Centralization}
As our experiment exhibits in Figure \ref{fig:mean_ratio}, the largest hubs in the network earn superlinearly more than nodes with lower capacity. This happens when the routing algorithm prioritizes routes with higher capacity than just fee minimization. Specifically, after running a new simulation for every value of \(\alpha\) and \(\mu\) we grouped the nodes in buckets of 200 elements. Then we computed the average ratio of each bucket. The buckets were created by decreasing capacity such that the first one contains the first 200 nodes in terms of capacity and the last one the last 200 nodes.
The bucketing procedure was employed to improve readability. From our results, many nodes across the 1000 considered routed 0 payments, as they were not in the shortest path so they were grouped together with those that effectively routed. The results with \(\mu = 0\) are not shown due to their high variance. When $\mu=0$, the sender considers the fee to pay to the routing node, and this results in a fee to pay some order of magnitude greater than the sending amount. The results of Figure \ref{fig:mean_ratio} show that there is an advantage in centralizing the liquidity in a single node. While for a high $\mu$ (or equivalently, low $\alpha$) the results for different buckets of nodes are almost the same, when $\mu$ decreases (or conversely $\alpha$ increases) the bucket of high capacity nodes shows a higher ratio than those of lower capacity.\\
In Table \ref{table:correlation} we illustrate how these results are correlated with the centrality measures of the nodes and with the node capacity. As we can see, while a higher correlation between the ratio and Betweenness Centrality is achieved by increasing $\mu$, it is also true that a decreasing value of $\mu$ increases the correlation between the ratio and node capacity.

\begin{table*}[t]
    \centering
    \caption{Correlation Between $Ratio(u)$ and Centrality Measures using Spearman correlation coefficient ( \(\rho = 1 - \frac{6 \sum d_i^2}{n(n^2 - 1)}
    \)) for values of $\mu$ with Capacity Proportional (CP) and Uniform Distribution of Payments}
    \label{tab:revised_centrality_correlation}
    
    \renewcommand{\arraystretch}{1.45}
    
    \begin{tabular}{|l|c|c|c|c|c|c|c|}
    \hline
     & \multicolumn{2}{c|}{$\mu=10$} & \multicolumn{2}{c|}{$\mu=100$} & \multicolumn{2}{c|}{$\mu=1000$} \\ \hline
    \textbf{Centrality Measure} & CP & Uniform & CP & Uniform & CP & Uniform \\ \hline
    
    Node Capacity: $\tau(u)$        & 0.53 & 0.54 & 0.47 & 0.46 & 0.44 & 0.39 \\ \hline
    Degree                          & 0.42 & 0.44 & 0.36 & 0.38 & 0.37 & 0.38 \\ \hline
    Closeness Centrality            & 0.36 & 0.38 & 0.30 & 0.31 & 0.33 & 0.33 \\ \hline
    Eigenvector Centrality          & 0.39 & 0.40 & 0.33 & 0.34 & 0.35 & 0.35 \\ \hline
    BC $w(e) = 1$                   & 0.43 & 0.47 & 0.37 & 0.41 & 0.39 & 0.40 \\ \hline
    BC $w(e) = fee_{ppm}$           & 0.35 & 0.38 & 0.46 & 0.53 & 0.62 & 0.66 \\ \hline
    \raisebox{0.5pt}{BC $w(e) = \frac{1}{c_e}$}        & 0.27 & 0.29 & 0.22 & 0.24 & 0.25 & 0.25 \\ \hline
    
    \raisebox{0.5pt}{BC $w(e) = \frac{15,000,000,000}{c_e} + \mu \cdot fee_{ppm}$} & 0.51 & 0.55 & 0.62 & 0.70 & 0.77 & 0.82 \\ \hline
    
    \end{tabular}
    \label{table:correlation}
\end{table*}

\section{Node strategies}
\label{section:nodes_rational_strategies}

The second aspect of our study concerns the rational behaviour of nodes. By analyzing the fee imposed by high capacity nodes it still results that many of these have little to no fee imposed \cite{beres2021cryptoeconomic}. Studies have been made on attachment strategies \cite{lange2021impact} for new joining nodes, but no previous research has explored how fee revenue is affected for nodes that are already active participants, nor has it examined the impact on the network when capacity becomes crucial for payment routing.

\subsection{Node replication}

The results of Section \ref{section:empirical_analysis} highlighted another relevant aspect of the high-capacity nodes we consider: there are many \textit{liquidity service providers} (LSPs) that currently are not earning routing fees and could benefit in terms of revenue by adopting more rational strategies in their channel selection and fee configuration. Furthermore, just in this subset of the network we count 3,468 edges with a final \(fee_e = 0\) on a total of 51,958 edges. Recent work has shown that \textit{Maximum Betweenness Improvement} (MBI) \cite{lange2021impact} is one of the best strategies to maximize a node's profit, but as showed in \cite{ersoy2020profit} selecting channels to maximize MBI is an NP-hard problem. Since \textit{Betweenness Centrality} (BC) computation for a given node \(u\) is instead of polynomial complexity, we illustrate that a node could increase its revenue by replicating the channels of other nodes with a better capital allocation and lower capacity. We say that node \(v\) has a better capital allocation than node \(u\) if node \(v\) channel selection and fee setting is such that \(R(v) > R(u)\). \\
In our analysis, we adopted two main strategies to illustrate that nodes can increase their profit with the same capital allocation.
\begin{itemize}
    \item \textbf{(1) Complete Information assumption:} 
    We assume that node \(u\) knows what is the node \(v\) with the best strategy (the highest ratio) to replicate. 
    Since the traffic is private, this cannot be computed exactly.
    However, in practice, a node should be able to approximate the answer via simulation.
    So through a simulation node \(u\) identifies node \(v\) and replicates $v$'s channel selection and policies after closing its channels. 
    \item \textbf{(2) Incomplete Information assumption:} this approach is typically employed by participants to maximize fee revenue under conditions of incomplete information. In this scenario, a player's knowledge of the graph topology is limited to what is publicly shared by the other nodes. The data presented in Table \ref{table:correlation} indicate that an increase in $\mu$ reduces the correlation between a node's revenue and its capacity. Conversely, as $\mu$ increases, the correlation with betweenness centrality strengthens. This supports previous findings that the optimal strategy for profit maximization involves achieving high betweenness centrality. To test this, we copied the strategy of nodes with the highest betweenness centrality in the network, assigning weights based on our greedy algorithm detailed in (1) (Section \ref{section:preliminaries}) and setting \(w(e) = \frac{15,000,000,000}{c_e} + \mu \cdot fee_{ppm}\).
    
\end{itemize}

It is important to note that, in this model, the cost associated with the opening and closing of channels is not taken into consideration in the analysis. Furthermore, the balance of the newly established channels is assumed to be split in half for each of the nodes participating in the channel. So, as it is allowed by the protocol we assume the channels are dual funded. In a real setting, to start earning payment fees it would be required to consider the cost of acquiring incoming liquidity.

\subsection{Opening and closing channel strategies}

When trying to replicate all the channels that a node already has, different possible strategies are available. Since the LN is a multigraph allowing multiple channels between nodes, when node \(u\) replicates node \(v\) channels, if node \(v\) has a channel \(e = (v,w)\) and there is already a channel \(e = (u, w)\) we can either create a completely new channel or close the already present channel \(e = (u, w)\) and create a new one such that \(c_e = c_{e_{uv}} + c_{e_{vw}}\). In the result, we assumed that a new channel is created, without exploiting the benefit a node could get with a new channel of increased capacity. \\
For the channel closure, we adopted a random approach. Every time we closed a channel we picked this channel uniformly at random. But a user could also adopt a more optimal strategy (e.g., close channels starting from the ones that routed less in the initial simulation).

\begin{figure*}[t]
    \centering
    \begin{minipage}{0.49\textwidth}
        \centering
        \includegraphics[width=\linewidth]{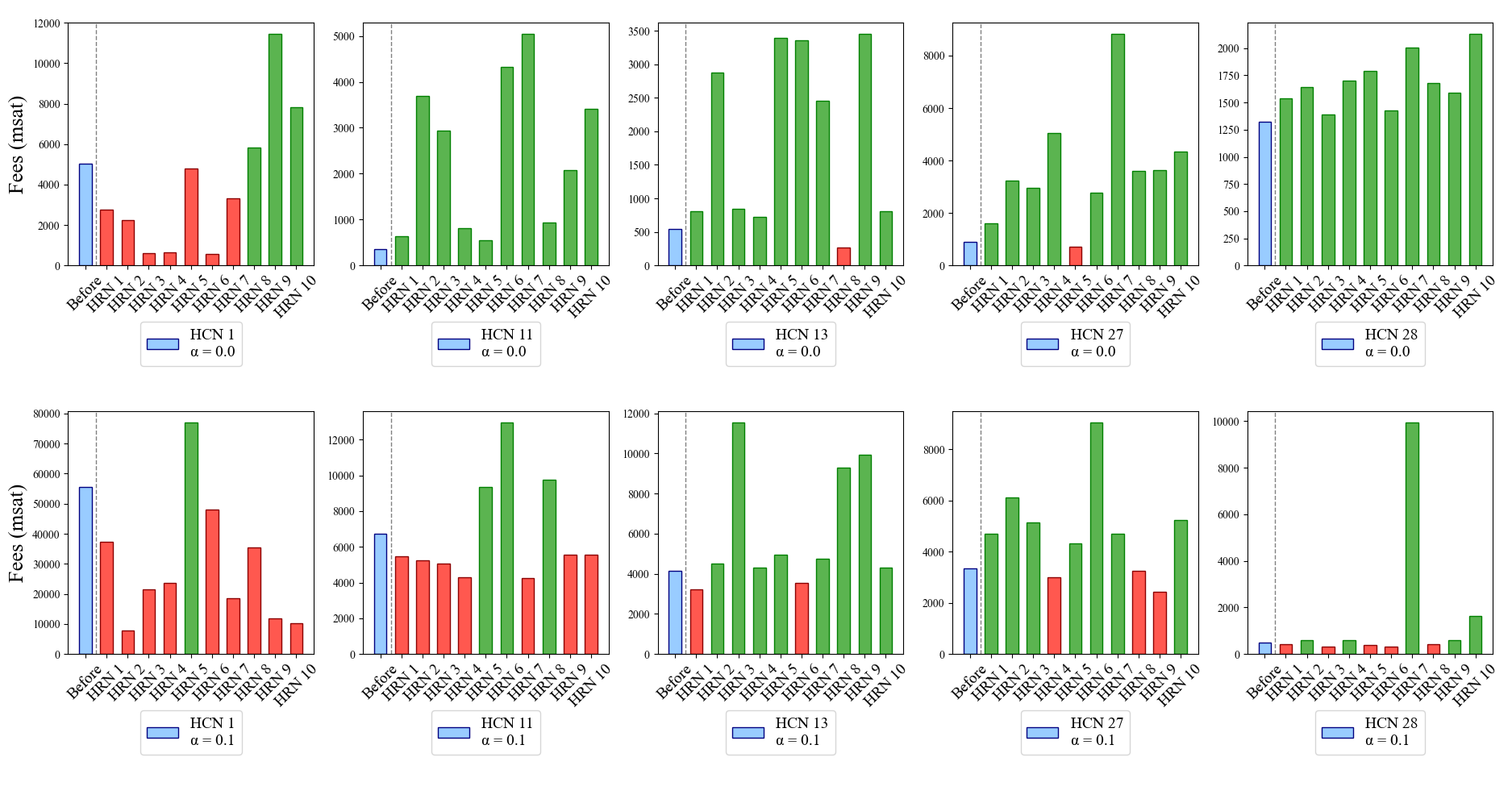}
        \caption{Fee revenue of 5 selected (HCN) before and after they replicate the first 10 highest Ratio Nodes (HRN) assuming a uniform distribution of payments in the network using our greedy routing algorithm for $\alpha= 0.0$  and $\alpha=0.1$ for simulations of 1,000 payments and 10,000 sats. Each simulation is independent from the others.}
        \label{figure:greedy_replication}
    \end{minipage}\hfill
    \begin{minipage}{0.49\textwidth}
        \centering
        \includegraphics[width=\linewidth]{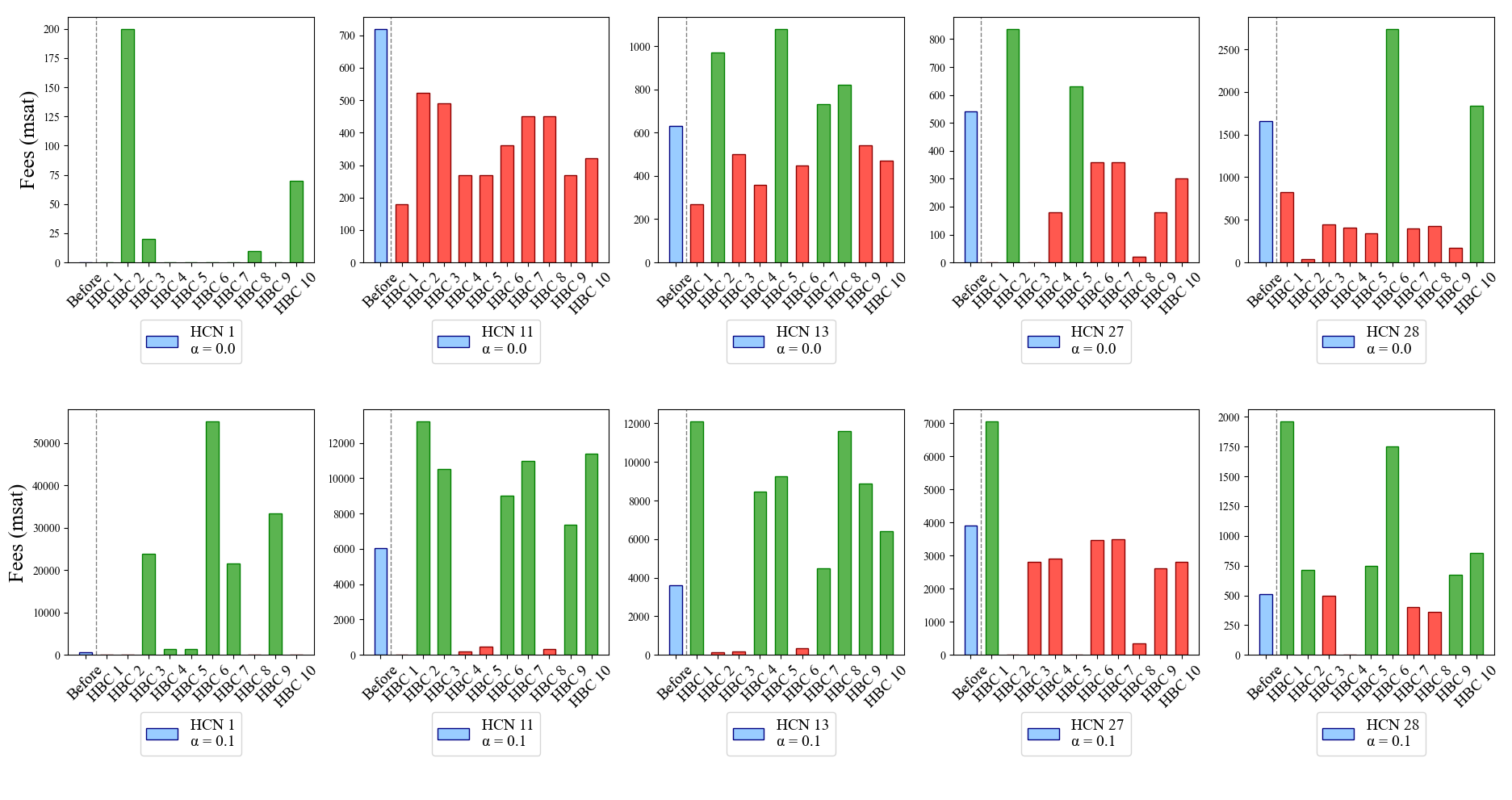}
        \caption{Fee revenue of 5 selected (HCN) before and after they replicate the first 10 Highest Betweenness Centrality Nodes (HBC) assuming a uniform distribution of payments in the network using our greedy routing algorithm for \(\alpha=0.0\) and \(\alpha=0.1\) for simulations of 1,000 payments and 10,000 sats. Each simulation is independent from the others.}
        \label{figure:pickhardt_replication}
    \end{minipage}
\end{figure*}

\subsection{Simulation setup}
The above strategies have been tested using the Pickhardt simulator and our simulator as described in Section \ref{section:empirical_analysis}, considering only successful payments of 10,000 sats. Differently from the previous section, the tests consisted of simulations of 1,000 payments. \\
In our analysis, we picked 5 of the nodes with the highest capacity \(\tau(u)\), specifically those already generating fees. This selection was based on the rationale that for nodes not on the shortest path of payments, thus not earning fees, the strategic change of strategy would trivially show an improvement. The second reason is to illustrate the impact on nodes already trying to maximize their routing income. The tests were performed with \(\alpha \in \{0.001, 0.1\}\) and \(\mu \in \{10, 1000\}\). As the first set of experiments in the previous section, the source and destination were selected only by uniform random sampling. Then, we picked 10 target nodes \(v\) to replicate according to the two different strategies (1) and (2). The results are shown Figure \ref{figure:greedy_replication} and \ref{figure:pickhardt_replication}. The replication strategy works as follows:

\begin{enumerate}
    \item Pick node \( u \in U \).
    \item Identify node \( v \in U \) to replicate such that \(\tau(u) > \tau(v)\).
    \item Close \( k \in \mathbb{N} \) channels of node \( u \) uniformly at random such that:
    \begin{center}
        \( \sum_{i=1}^{k} b(e_{u_{i}}) \geq \tau(v) \)
    \end{center}
    
    \item Replicate the channels of node \(v\) such that \(N(u) = N(v)\). For every edge \(e_{uw}\) connecting \(u\) to a neighbor \(w \in N(u)\) and the corresponding edge \(e_{vw}\) connecting \(v\) to \(w\) the new channels are such that:
    \begin{center}
        \(c(e_{uw}) = c(e_{vw}) + \epsilon\)
    \end{center}
    with $\epsilon > 0 \in \mathbb{N}  $
     The same procedure is applied to the fees if they have values greater than \(0\):
    \begin{center}
        \(fee_{ppm}(e_{uw}) = fee_{ppm}(e_{vw}) - \zeta\) \\
        \(fee_{base}(e_{uw}) = fee_{base}(e_{vw}) - \iota\) \\
    \end{center}
    with $\zeta > 0 \in \mathbb{N}$ and $\iota > 0 \in \mathbb{N}$.
    
\end{enumerate}

We maintained the same fee of the edge replicated in the opposite direction.

\subsection{Results and discussion}

In Figure \ref{figure:greedy_replication} and \ref{figure:pickhardt_replication} we illustrate the results of our analysis. 
For each of the HCN (high capacity node) selected we show the fee revenue in the initial condition (in blue) compared to the fee revenue when they re-allocate their capital by replicating the first 10 highest ratio nodes (HRN) and the first 10 highest betweenness centrality nodes (HBC) in the network having less capacity than them. The general trend is that the strategies adopted are effective and nodes do increase their profit in the majority of their replications. However, not all replication efforts lead to an increase in profits. This is influenced by the chosen strategy for closing channels. In certain situations, such as HCN 1 in Figure \ref{figure:greedy_replication}, the existing allocation of capital is sufficiently effective, thereby diminishing the potential benefits of strategic modifications.

\begin{table}[h]
    \centering
    \caption{Alias of the Highest Capacity Nodes (HCN)}
    \begin{tabular}{|c|c|}
    \hline
    \textbf{(HCN) Rank} & \textbf{Node Alias}                                       \\ \hline
    1                                            & bfx-lnd0                         \\ \hline
    11                                           & LNBiG.com [lnd-19/old-lnd-11]    \\ \hline
    13                                           & LNBIG.com [lnd-49/old-lnd-13]    \\ \hline
    27                                           & LNBiG.com[lnd-20]                \\ \hline
    28                                           & Boltz                            \\ \hline
    \end{tabular}
    \label{tab:node_capacity}
\end{table}

\section{Long-term impact of random strategies}
\label{section:long_term_impact_random_strategy}

In the concluding section of this study, we turn our attention to the consequences of a scenario in which all the nodes base their decisions on rationality from an economic perspective. So, we investigated network stability over time. As suggested in \cite{avarikioti2020ride}, at every step of the iteration, nodes look at their current revenue, and if they obtain an increase in profit, they keep the new configuration. Otherwise the revert back to the old one.

\subsection{Experimental Methodology}
The setting is similar to the one in previous sections. We simulated 10,000 payments of 10,000 sats. As computing the shortest path is a computationally expensive task, we pre-compute all shortest path pairs in the network before running the simulation, since the complexity is \(O(V  E\log V)\).\\
We implemented a methodology consisting of perturbing all the nodes in the snapshot. In the first experiment, we change the fee of one channel for all the nodes, while in the second experiment we change one neighbor for all the nodes in the network.

We impose an upper bound of 500 msat on the fee paid for each payment, since for a fee greater than this amount the end user will likely switch to a payment method outside the LN.\\
We then simulate successful payments for which a payment path is available. Since the 500 msat threshold is imposed, the randomly chosen final \(fee_e\) for the edges (that takes into consideration both \(fee_{base}\) and \(fee_{ppm}\)) is imposed in the interval \([1, 500]\) msat.

\subsection{Random fee}
In this experiment, we set a new fee to the selected node channels, while maintaining the same neighbors. 
The fee of the new channel was chosen with a value that is obtained through a discretized exponential distribution. Specifically, the new fee (that considers both $fee_{base}$ and $fee_{ppm}$ ) of node $u$ to node $v$, is determined by the expression $fee(e_{uv}) = \left\lceil -\frac{\ln(1 - U)}{\lambda} \right\rceil$, where \( U \) is a uniformly distributed random variable in the interval \( (0, 1) \), and \( \lambda \) is the rate parameter of the exponential distribution, defined as \( \lambda = \frac{1}{\text{scale}} \), with the scale being a predetermined positive real number (in our case \(scale=50\)). The resulting $fee$ is constrained to the inclusive range $[1, L]$ with $L = 250$.

\begin{figure*}[t]
    \centering
    \begin{minipage}{0.49\textwidth}
        \centering
        \includegraphics[width=\linewidth]{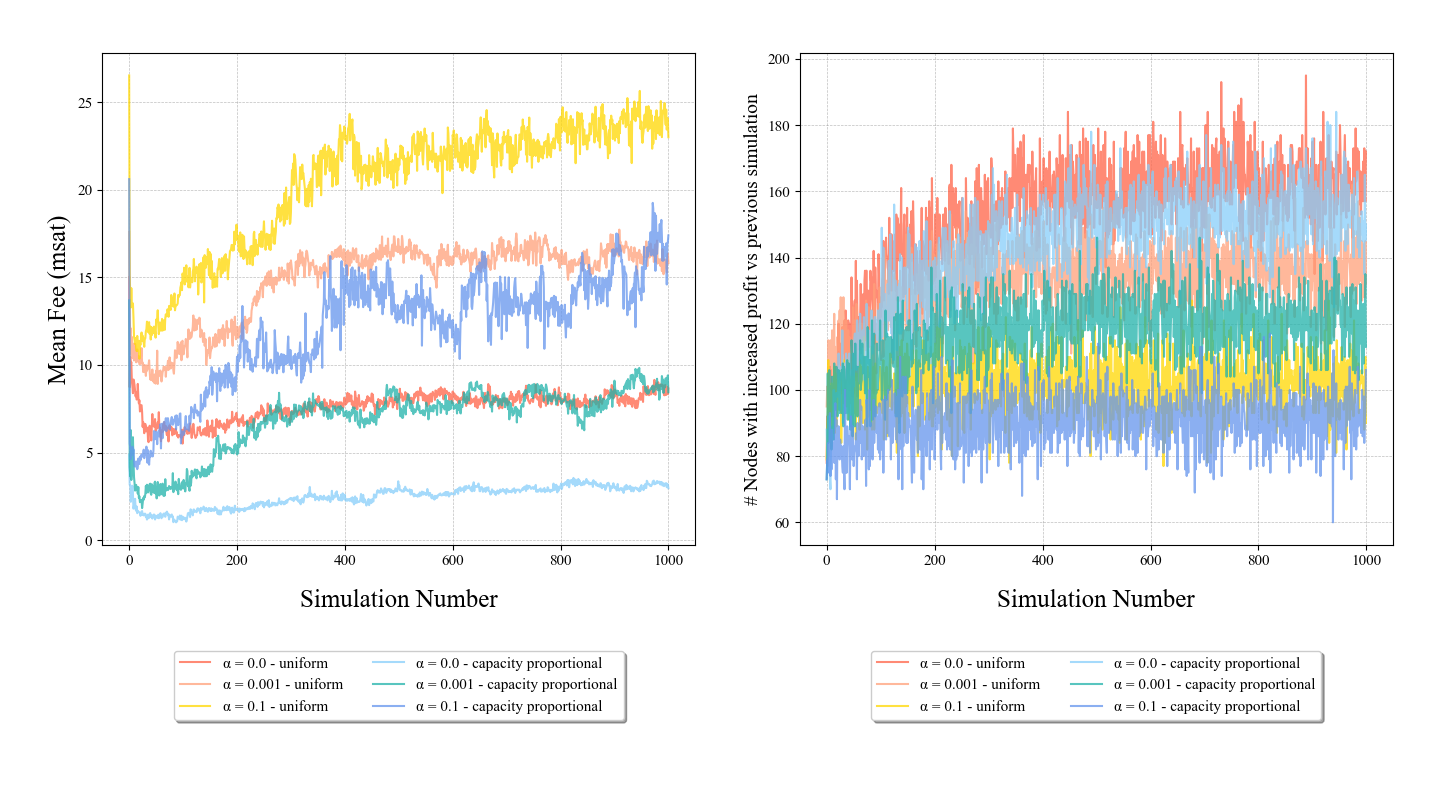}
        \caption{Average fee paid per simulation when nodes can only adjust their fees.}
        \label{figure:experiment3_change_fees}
    \end{minipage}\hfill
    \begin{minipage}{0.49\textwidth}
        \centering
        \includegraphics[width=\linewidth]{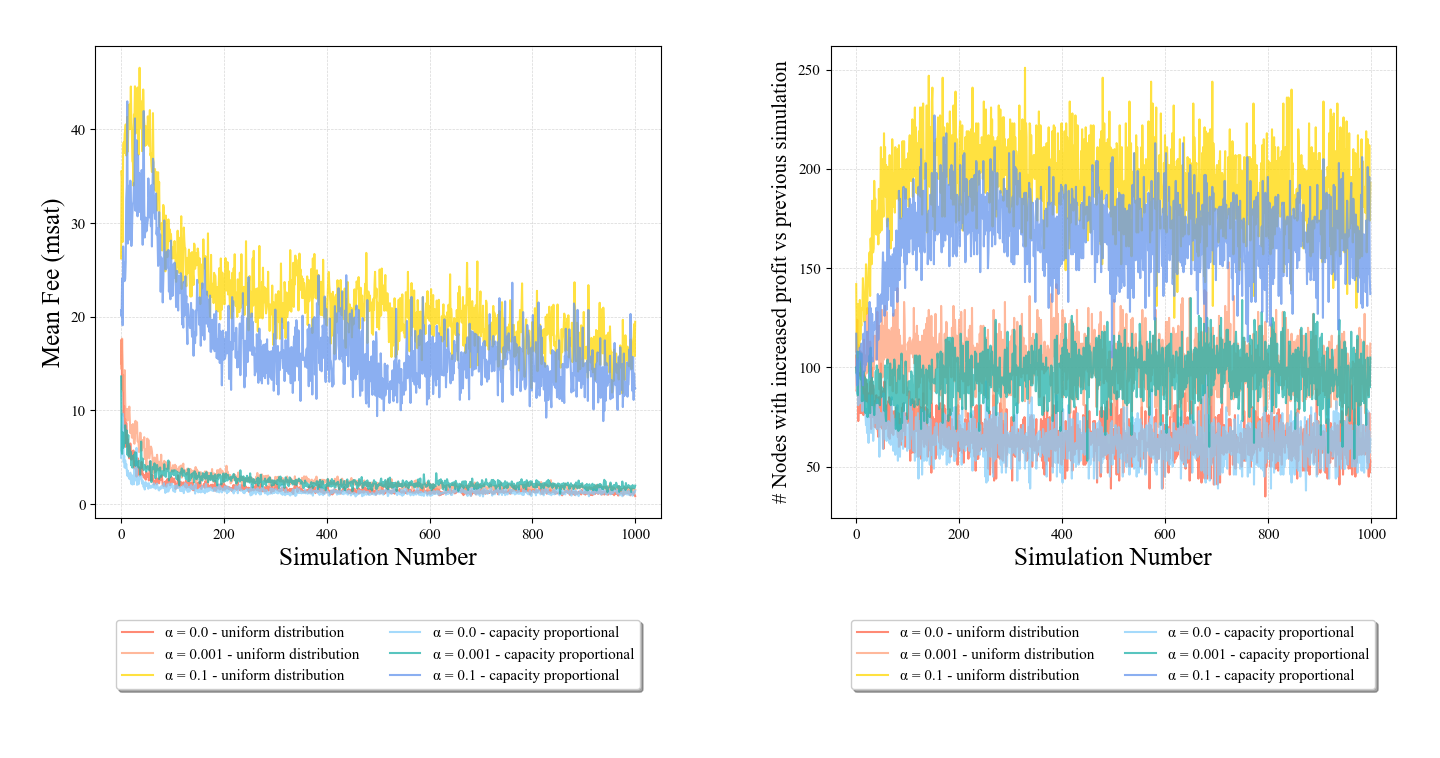}
        \caption{Average fee paid per simulation when nodes can change only their neighbors.}
        \label{figure:experiment3_change_channels}
    \end{minipage}
\end{figure*}

\subsection{Random channels - single funded}
In this series of experiments, we altered the neighbors of the selected nodes instead of the fee policy. Whenever a channel was closed, a new channel was established by selecting a new neighbor for the node uniformly at random. To ensure consistent network capacity across successive simulation rounds, we implemented a capacity redistribution mechanism. Specifically, when a channel between nodes \( u \) and \( v \) was closed, half of its capacity was reallocated among \( v \) existing channels, denoted as \( N(v) \). This redistribution was based on the assumption that only channels connected to nodes with at least one remaining channel were closed. The redistributed capacity for each of \( v \)'s remaining channels was calculated as \( c_{vw}' = c_{vw} + \frac{c_{uv}}{2 \cdot deg^+(v)} \), where \( deg^+(v) \) represents the number of channels associated with node \( v \), for simplicity. This resizing of node $v$'s channels was conducted assuming that new protocol updates will allow this procedure. The other half of the capacity from the closed channel was used to form a new, single-funded channel with a randomly chosen node \( z \).  The fee for this new channel was determined by taking the median of the fees imposed by \( z \) in its other channels.

\subsection{Results and discussion}
\begin{figure*}[t]
    \centering
    \includegraphics[width=\textwidth]{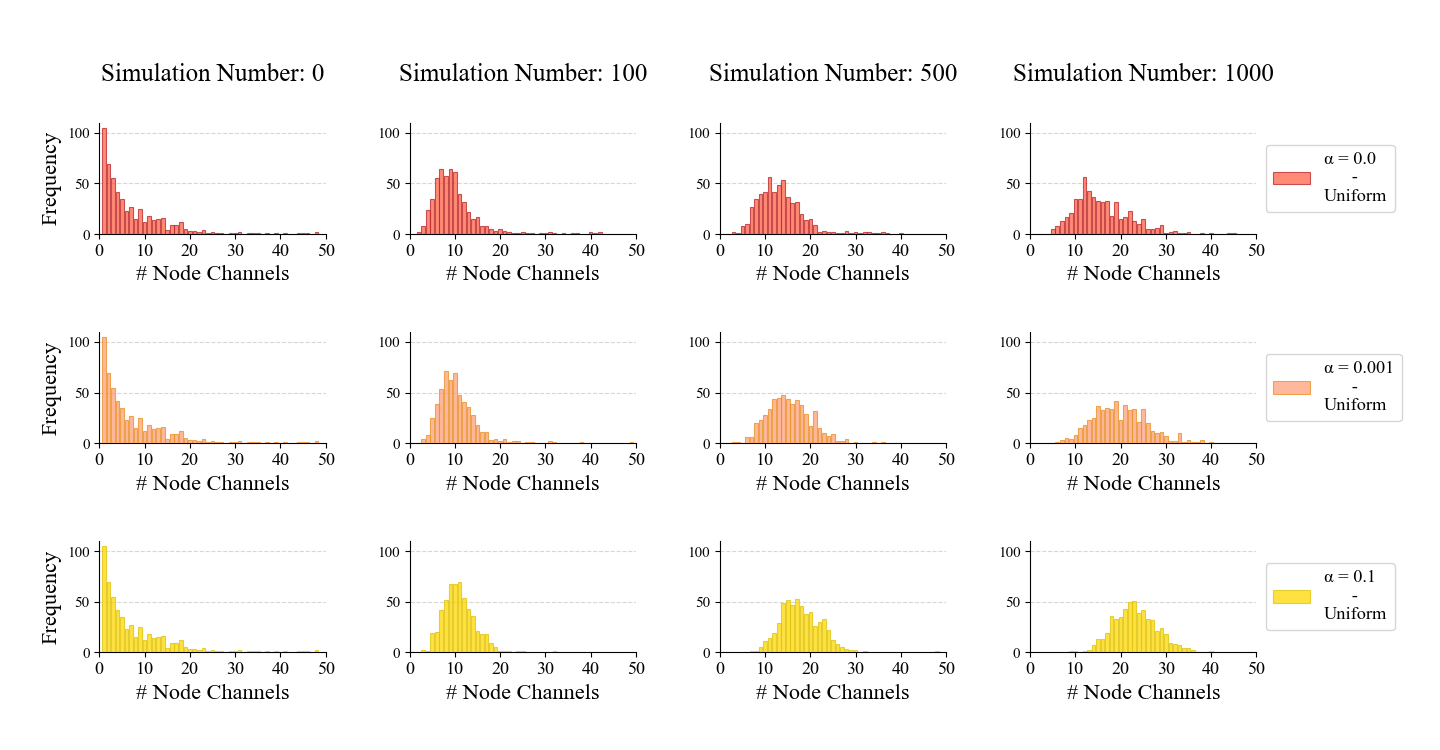}
    \caption{Nodes' channels distribution in selected simulations over 1000 simulations of 10,000 payments with the random channels strategy.}
    \label{figure:new_degree_distribution}
\end{figure*}

The results are shown in Figures \ref{figure:experiment3_change_fees} and \ref{figure:experiment3_change_channels}.
To evaluate the outcomes of the random fee experiment, we monitored the average fee paid by participants in the network over 1000 simulations. This analysis aimed to understand the impact of varying levels of $\alpha$, particularly its influence in favoring channels with larger capacities. Our findings were consistent across the two different payment distribution methods tested.
We observed a sharp decrease in the mean fee as the simulations begin, as a consequence of the imposition of the new fee policy for all the nodes in the network (in the experiments where the new fee policy was applied to a smaller number of connections, the initial drops in the average fee were less pronounced). Then, up to the 400th round of simulation, there was a marked rise in the mean fee with higher $\alpha$ values. Beyond this point, until the 1000th round, the rate of increase in the mean fee stabilized. An observation from our analysis is that network participants tend to benefit more from lower $\alpha$ values, where the mean fee is lower and increases at a lower rate. This pattern suggests that in scenarios where nodes are restricted from changing their network connections, nodes with larger capacity channels leverage their bottleneck position to form coalitions more effectively.\\
In contrast, when examining scenarios where channel connections are changed, the dynamics differ significantly.
The competitive nature among nodes in this scenario leads to a reduction in the mean fee as each node aims to increase its profits.\\
Another important result of the strategy of changing only nodes' channels is the change in the degree distribution over time with this strategy, shown in Figure \ref{figure:new_degree_distribution}. While the probability of an edge being removed from a node is proportional to its degree, random reassignment of neighbors suggests that the network would retain its initial degree distribution if it were initially in equilibrium. Our findings show that this is not the case, and even if the future network development will probably not exactly follow our assumptions as new nodes continue to join the network, these results provide a scenario where nodes with high capacity dictate network evolution. 
We thus observe that, over time, the degree distribution of the highest capacity nodes tends to converge to a uniform distribution.

\section{Related work}
\label{section:related_work}

Previous works have investigated the economic behaviour of nodes in the LN \cite{beres2021cryptoeconomic}, with also a focus on privacy concerns due to the LN small world nature. While game theoretic studies investigated graphical games from a theoretic standpoint \cite{yau2021approximating, avarikioti2020ride} and proved that centralized structures can make the network efficient, evaluating these studies empirically has been proved to be challenging.\\
Instead, the field of centrality measures to analyze the influence of nodes is prolific \cite{singh2020node, bloch2023centrality, epstein2009efficient, puzis2012heuristics}.\\
In our work, we empirically evaluated the best attachment strategies that from other studies emerged as the most effective \cite{lange2021impact}. Also Ersoy, Roos and Erki in \cite{ersoy2020profit} proved that improving the betweenness results in the increase of a node's fee revenue.\\
Finally, our study takes into account the consequences of using a routing algorithm that prioritizes high capacity channels as introduced by the recent work of Pickhardt and Richter in \cite{pickhardt2021optimally}. While they showed that prioritizing high capacity channels increases payment success rate, we pose the question of how this would affect the network and whether this would lead to a more centralized topology. To this end, Avarikioti, Heimbach, Wang and Wattenhofer in \cite{avarikioti2020ride} follow a game-theoretic approach and show that centralized structures can make the network more efficient and stable. As a consequence, our work puts itself on the line of the fact that attachment strategies optimizing for profit and efficiency tend to favor the creation of highly centralized topologies, highlighting the results of the contradictory trade-off between network efficiency and decentralization.

\section{Conclusion}
\label{section:conclusion}

In conclusion, this work highlighted the \textit{``rich-get-richer"} phenomenon within the Bitcoin Lightning Network. We empirically analyzed the routing income of high-capacity nodes and found that these nodes tend to accumulate fees more rapidly, further cementing their dominant positions. Moreover, our findings raise questions about how to address challenges in achieving fee distribution equity.\\
Then, our analysis revealed that numerous nodes, including some of the most influential ones, have not yet implemented a fee policy that is optimized for maximizing earnings from routing fees, therefore we suggested simple techniques to achieve this objective. \\
While it is important to highlight that network participation could be beyond the immediate financial gains from routing fees, the strategic actions and fee policies of influential nodes are crucial in dictating the network's economic landscape. Further exploration into strategies that can counterbalance the growing disparity and ensure a more equitable fee structure is still required.\\
To conclude, our latest findings highlight that, if a time of high on-chain fees occurs, this routing structure could lead nodes to limit their channel openings and neighbors change and, as a consequence, higher fees paid by users. Instead, in the unlikely scenario where on-chain fees will decrease, competition will favor users against nodes that are highly central in the network. 

\bibliographystyle{IEEEtran}
\bibliography{bibtex}

\end{document}